\documentclass[%
 reprint,
 amsmath,amssymb,
 aps,
10pt,
]{revtex4-2}

\usepackage{graphicx}
\usepackage{dcolumn}
\usepackage{bm}
\usepackage{xcolor}
\usepackage{comment}
\usepackage{subcaption}
\usepackage{soul}
\usepackage[normalem]{ulem}

\newcommand{\V}[1]{\ensuremath{\boldsymbol{ #1}}}

\newcommand{\Vh}[1]{\ensuremath{\hat{\boldsymbol{ #1}}}}

\newcommand{\del}{\V{\nabla}}

\newcommand{\pd}[2]{\frac{\partial #1}{\partial #2}}

\newcommand{\eqnref}[1]{eqn.~\ref{eqn:#1}}
\newcommand{\Eqnref}[1]{Eqn.~\ref{eqn:#1}}
\newcommand{\figref}[1]{Fig.~\ref{fig:#1}}

\newcommand{\Sectionref}[1]{Section \ref{sec:#1}}

\newcommand{\tableref}[1]{Table \ref{table:#1}}

\newcommand{\appendixref}[1]{Appendix \ref{sec:#1}}

\begin{document}


\title{Stretching Behavior of Knotted and Unknotted Flow Fields}

\author{Stefan Faaland}

\author{Diego Tapia Silva}

\author{Dustin Kleckner}%
\email{dkleckner@ucmerced.edu}
\affiliation{ University of California, Merced\\
}

\date{\today}

\begin{abstract}
Vortex stretching is a common feature of many complex flows, including turbulence.
Experiments and simulations of isolated vortex knots demonstrate that this behavior can also be seen in relatively simple systems, and appears to be dependent on vortex topology.
Here we simulate the advection of material lines in the frozen flow fields of vortices on the surface of a torus.
We find that knotted configurations lead to exponential stretching behavior which is qualitatively different than that observed by collections of unknots.
This stretching can be explained by the formation of bights, sharp bends in the material lines which can be used to predict the stretching rate. 
This behavior is confirmed by computing the finite time Lyapunov exponents of the flow fields, which demonstrate the exponential stretching is mediated by bight forming regions between the vortex lines.
This work both establishes a clear connection between topology and stretching behavior, as well as providing an intuitive mechanism for exponential growth of material lines in knotted flows.

\end{abstract}
\maketitle

\section{\label{sec:level1}Introduction:}

\begin{figure}
    \includegraphics{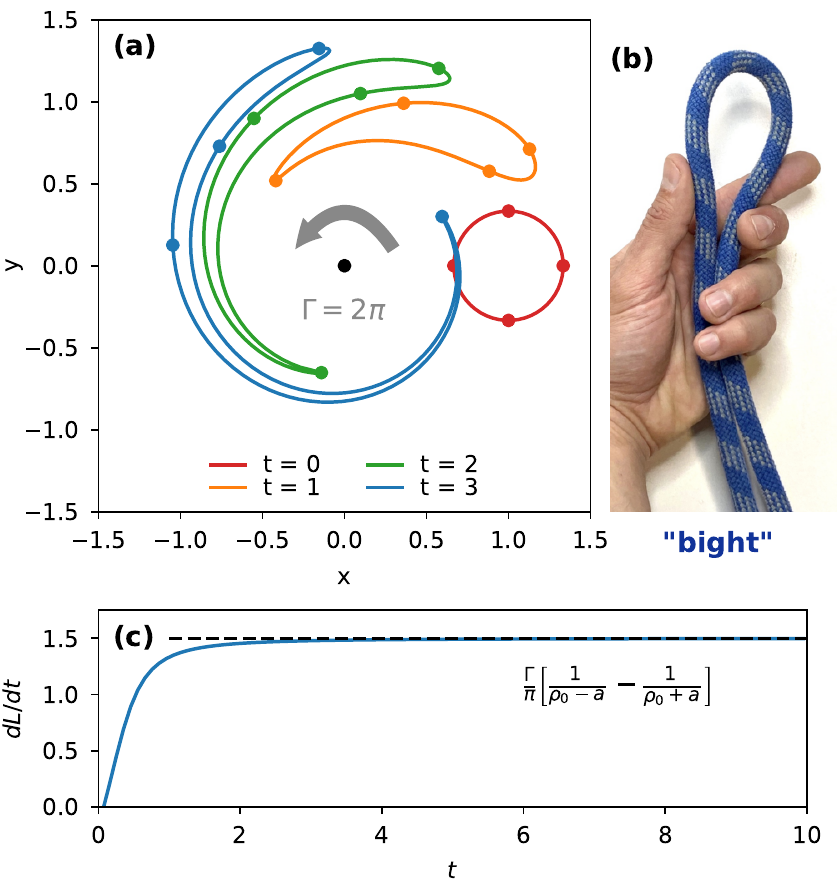}
    \caption{
        \textbf{(a)} The time evolution of a circular material loop (radius $a=1/3$, and initially centered at $\rho_0 = 1$) advected in the flow field of an infinite line vortex with circulation $\Gamma =2 \pi$, oriented along $\Vh z$.  
        As the initially circular loop evolves, it develops compact `bights' at the leading and trailing edges.
        \textbf{(b)} A bight in a physical piece of rope, as used in knot tying.
        \textbf{(c)} The stretching rate of the material line, $dL/dt$.
        This rate quickly settles to a constant value, which can be predicted using only the velocity field calculated at the locations of the two `bights' which form as the loop is stretched.
        \label{fig:bight}
    }
\end{figure}

There are many examples of flows with concentrated vortex lines, including tornadoes \cite{snowReviewRecentAdvances1982}, smoke rings  \cite{brunelliVanishingViscosityLimit2011}, flows inside the heart 
\cite{kheradvarVortexFormationHeart2012}, and turbulence \cite{douadyDirectObservationIntermittency1991, barenghiSuperfluidVortexLines1997}.
Indeed, any incompressible flow can be regarded as a collection of vortex lines by the Helmoltz Theorem, and so understanding their behavior provides a complete description of such flows.
As a result, there has been considerable theoretical, numerical, and experimental research focused on understanding the behavior of concentrated vortex filaments (e.g.~\cite{kleinSimplifiedEquationsInteraction1995, bialynicki-birulaMotionVortexLines2000, bedrossianVortexFilamentSolutions2020, leonardNumericalSimulationInteracting1975, pumirNumericalSimulationInteracting1987, mcgavinVortexLineTopology2018, limExperimentalStudyVortex1989, kwonExperimentalStudyVortex1989, vargaDynamicsDensityQuantized2018}).

While a simple circular vortex ring is known to be quite stable \cite{maxworthyStructureStabilityVortex1972, krutzschExperimentallyObservedPhenomenon2011}  -- even if distorted \cite{kambeMotionDistortedVortex1971, krutzschExperimentallyObservedPhenomenon2011} -- knotted and tangled vortices have been observed to be highly unstable in simulations \cite{klecknerHowSuperfluidVortex2016,binyshStableUnstableVortex2019, kimuraScalingPropertiesVortex2018, kerrTrefoilKnotTimescales2018, promentTorusQuantumVortex2014} and experiments \cite{klecknerCreationDynamicsKnotted2013, klecknerLifeVortexKnot2014}.
In particular, tangled vortex lines are observed to rapidly stretch, leading to vortex reconnections which ultimately untie and/or dissipate the flow \cite{klecknerCreationDynamicsKnotted2013}.
This behavior is quite reminiscent of the features of turbulent flows \cite{buariaVortexStretchingEnstrophy2020,douadyDirectObservationIntermittency1991,moffattStretchedVorticesSinews1994,pullinVortexDynamicsTurbulence1998}, and has both enstrophy production (vortex stretching) and transport of energy to small scales, where it is dissipated (vortex reconnections).


The connection between vortex stretching and reconnections can be understood in terms of energy conservation: in order to stretch vortex lines without increasing energy, it is necessary to create regions of closely spaced counter-rotating vortices \cite{klecknerHowSuperfluidVortex2016, kerrTrefoilKnotStructure2017}.
As the vortices continue to stretch, these counter-rotating vortices must get closer together, ultimately resulting in reconnections which continue until non-stretching vortex state is reached \cite{klecknerCreationDynamicsKnotted2013, klecknerHowSuperfluidVortex2016}.
Moreover, vortex stretching (i.e.~enstrophy production) is a key feature of turbulence; indeed it can be regarded as the key feature which separates 2D from 3D turbulence \cite{ouelletteTurbulenceTwoDimensions2012}.
As a result, vortex stretching is intimately connected to the stability of flows, motivating research in to how and why collections of vortex lines self-stretch.

Previous work has modeled this stretching using either vortex filament models \cite{kleinSelfstretchingPerturbedVortex1991, kimuraTentModelVortex2018} or direct numerical simulations of Newtonian or super-fluids using the Navier-Stokes \cite{bedrossianVortexFilamentSolutions2020} or Gross-Pitaevskii equations \cite{villoisEvolutionSuperfluidVortex2016}.
In either case, the non-linearity of fluid flows makes these models difficult to accurately simulate, and complicates the interpretation of the results.

Here we take a different approach: `freezing' vortex generated flow fields, and investigating how material lines stretch when advected in this flow.
Given that vortex lines themselves are transported by the flow \cite{katopodesVorticityDynamics2019}, it follows that a flow field which stretches material lines will also stretch vortices.
In this manuscript, we investigate if this stretching can be understood in terms of the properties of the flow field at a single instant in time, rather than as a consequence of the non-linear evolution of the flow.
Using this approach, we find that simple unlinked and unknotted vortices have a flow field which produces linear stretching, while the flow field of knotted vortices produces regions of exponential stretching.
Moreover, this stretching can be attributed to the generation of `bights' in the material lines (\figref{bight}), and that exponential stretching is only possible if these bights are continuously produced.
This result offers a potential connection to existing results in two-dimensional topological mixing \cite{boylandTopologicalFluidMechanics2000}, and appears to explain why knotted vortices are themselves unstable.
Moreover, the concept of bights provides a simple mechanism for understanding how lines in vortex dominated flow fields should stretch over time, with potential applications to a variety of fluid flow problems.

\section{Stretching from a Line Vortex}
\label{sec:inf-bights}
Before discussing the stretching produced by the flow fields of complex vortex geometries, it is useful to consider the case of a straight line vortex.
The flow field of an infinite vortex along with the $z$-axis is given by:
\begin{align}
	\V u &= \frac{\Gamma}{2\pi \rho}\Vh \phi,
\end{align}
where $\Gamma$ is the circulation, $\rho$ and $\phi$ are cylindrical coordinates defined in the usual way, and $\Vh \phi$ is a unit vector in the azimuthal direction.
Here, and in the rest of this manuscript, we will use dimensionless spatial and time coordinates, and a dimensionless circulation of $\Gamma = 2 \pi$ (i.e.~for an infinite straight line we obtain: $\V u = \Vh{\phi}/\rho$).

The evolution of a small circular material line in this flow field is shown in \figref{bight}.
How can we understand the stretching of this line?
The time derivative of the total length of the material line, $L$, is given by:
\begin{align}
    \dot{L} &= \oint  \underbrace{\del_{\Vh T} \V u}_{\frac{\partial \V u[\V r(s)]}{\partial s}} \cdot \Vh T\ ds\\
    &= -\oint \V u \cdot \kappa \Vh N\ ds, 
    \label{eqn:curve-stretch}
\end{align}
where $\V r(s)$ is the material line displacement as a function of arc length coordinate, $s$.
$\Vh T$, $\Vh N$, and $\kappa$ are the Frenet-Serret tangent vector, normal vector, and curvature, respectively, which obey the relationships $\Vh T = \pd{\V r}{s}$ and $\kappa \Vh N = \pd{\Vh T}{s}$.
\Eqnref{curve-stretch} is obtained using integration by parts, and would have an additional term for an open material line.
(Note that integrand of \eqnref{curve-stretch} does not give the \emph{local} stretching rate; see \appendixref{global-local}.)

Any section of material line advected in the flow of an infinite line vortex will tend to align or anti-align with $\V u$ as $t \rightarrow 0$
(e.g.~\figref{bight}; see also supplemental materials for a proof).
When this happens, $ \V u \cdot \Vh N \rightarrow 0$,
and from \eqnref{curve-stretch} we would expect that these regions to produce no stretching.

A closed material line, however, must change direction with respect to $\V u$ at two or more points to form a closed path.
Over time, these regions will form compact 180$^\circ$ bends, which we will refer to as `bights', by analogy with a term using in knot tying.
Formally, we will define a bight as the point at which $\V u \cdot \Vh T$ changes sign along a material line.

Near the bight, $ \V u \cdot \Vh N \neq 0$, and so the bights are responsible for stretching or shortening the material line, depending on whether they are a `leading' bight ($\Vh N \cdot \Vh u < 0$) or a `trailing' bight ($\Vh N \cdot \Vh u > 0$).
(Note that for \emph{open} material lines, the ends of the material line also function similar to bights; see Supplemental Materials for details.)

If we assume that \emph{all} the stretching can be attributed to bights -- each of which is assumed to be a compact 180$^\circ$ bend -- we would obtain a simple expression for the stretching rate:
\begin{align}
    \label{eqn:bight-stretch}
   \dot L(t \rightarrow \infty)  &\approx -2 \sum_{\rm bights} \Vh N \cdot \V{u},
\end{align}
where $\V u$ and $\Vh N$ are computed at the bights and an overall factor of 2 is obtained by integrating $\V u \cdot \kappa \Vh N$ around the 180$^\circ$ bend.
As an example, consider a material line which is a circle displaced from a single vortex line (\figref{bight}).
If this circle has center displacement, $\rho_0$, and radius, $a$, it will form a leading bight near the location closest to the vortex, at $\rho = \rho_0 - a$, and a trailing bight near the spot furthest from the vortex, at $\rho = \rho_0 + a$.
Under the assumption that each bight is a 180$^\circ$ bend in the material line, we would predict a stretching rate of:
\begin{align}
    \dot L(t \rightarrow \infty) = \frac{\Gamma}{\pi} \left[\frac{1}{\rho_0 - a} - \frac{1}{\rho_0 + a}\right].
\end{align}
As can be seen in \figref{bight}\textbf{c}, the stretching rate converges on this result after only a fraction of a turn around the vortex.

This behavior is not unique to circular material lines: as shown in the supplementary materials, \emph{any} material line should tend to align with the flow field of a straight line vortex over time.
As a result, it will form some number of compact bights, and in the long time limit these will determine the stretching rate.

\section{Stretching from Vortex Rings and Knots}

It is not clear that such a formulation should apply to more complicated geometries, for which we can not assume that material lines will always align with $\V u$.
In this section we consider the stretching of material lines in the flow fields of more complex vortex shapes, including rings, distorted rings, and knots.

\begin{table*}
\begin{tabular} { c|c|c|c|c } 
 \textbf{Vortex Shape} & \textbf{Aspect Ratio} & \textbf{Tor.~Rate} & \textbf{Pol.~ML Radius} & \textbf{Pol.~Rate}\\
 \hline
& $a=0.1$ & $k_t = 31.5$ & $r_p = 0.15$ & $k_p = 27.9$\\
\cline{2-5}
 & $a=0.2$ & $k_t = 5.30$ & $r_p = 0.3$ & $k_p = 5.08$\\
\cline{2-5}
 & & & $r_p = 0.35$ & $k_p = 2.16$\\ 
 &&& $r_p = 0.45$ & $k_p = 2.36$\\ 
Trefoil Knots &$a=0.3$ & $k_t = 2.48$& $r_p = 0.55$ & $k_p = 2.37$\\ 
 &&& $r_p = 0.65$ & $k_p = 2.25$\\ 
 &&& $r_p = 0.75$ & $(k_p = 0.237)^{\dagger}$\\  
 \cline{2-5}
& $a=0.4$ & $k_t = 1.41$ & $r_p = 0.6$ & $k_p = 1.41$
\end{tabular}
\caption{
Measured exponential stretching rates for torus knots.  In each case, an exponential stretching rate, $k$, is determined by a least squares fit of $\ln L(t) = a + k t$.
To eliminate initial transients, the fit is over only the last 1.5 time units of each simulation. 
Fits are shown for initial material lines that are both toroidally (tor.) and poloidally oriented (pol.).  
In the latter case the radius of the initial material line, $r_p$ is indicated.
$\dagger$ indicates data sets which do not fit well to exponential growth curves; see \figref{stretch} and \figref{pol-stretch}. 
}
\label{table:exp}
\end{table*}

\subsection{Methods}

The flow field of more complex vortex shapes is computed using the Biot-Savart law:
\begin{equation}
    \V u(\V x) = \frac{\Gamma}{4\pi} \oint \frac{\Vh T \times (\V{r} - \V x)}{|\V r - \V x|^3} ds,
\end{equation}
where here $\V r(s)$, $\Vh T$, and s refer to the vortex path (rather than a material line), and we set $\Gamma = 2 \pi$ as before.
In cases where there is more than one vortex, the velocities from each are summed.
In practice, the vortex path(s) are represented as polygons with a total of 100 points, in which case an exact expression can be obtained for the flow field \cite{hansonCompactExpressionsBiot2002}.
Because we are considering frozen flow fields, we do not advect the vortex paths in time -- as would happen in flow described by the Navier-Stokes equation -- but rather treat them as fixed.

The strain rate tensor of this field ($\del \V u$) can be explicitly computed, allowing for the advection of infinitesimal vectors attached to each points.
This allows us to represent material lines as piece-wise Bezi\'er curves whose control points are computed using a tangent vector attached to each end point.
These end points and vectors are integrated in time using the Dormand-Prince method \cite{dormandFamilyEmbeddedRungeKutta1980, shampinePracticalRungeKuttaFormulas1986} with an absolute velocity tolerance of $10^{-8}$.
(See \appendixref{sim-details} for more details on our numerical scheme.)

To predict stretching from bights, we identify the locations on the material line where $\Vh T \cdot \V u$ changes sign, and then compute the normal vector from the implicit Bezi\'er curve.
This allows us to directly compute \eqnref{bight-stretch}, and compare this to the actual stretching rate found by the derivative of the length of the material line.
Note that, in the case of the more complicated geometries discussed below, we may also obtain bights for relatively straight sections of the material line where the flow field changes direction, rather than being caused by a tight bend in the material line.
These can be filtered by curvature or other means, however, this does not significantly affect our results and so we neglect it in the following discussion.
(In practice, we have observed that these `phantom' bights are typically a small fraction of the total bights, and often become `true' bights as they evolve in time.)

In general, each of the simulations is run until either the number of resolved points exceeds $5 \times 10^7$, or the simulation fails due to lack of precision.
The latter condition typically happens when the bights themselves become sufficiently sharp that the can not be properly resolved by double precision floating point numbers.

\subsection{Role of Geometry and Topology in Stretching of Material Lines}
In order to understand how shape and topology affect the stretching behavior of 3D vortices, we will consider several vortex shapes: pairs of circular or distorted rings and trefoil knots of varying aspect ratio.
To allow for better comparison between different cases, each of the vortex lines is defined on the surface of a torus with major radius, $R=1$, and minor radius (and aspect ratio), $a = 0.1-0.4$.
Four different aspect ratios are used in the case of the knotted vortices, while $a=0.3$ is used for all unknotted vortices.
The parametric equations for all vortex lines are given in \appendixref{sim-params}.

\begin{figure*}
\includegraphics{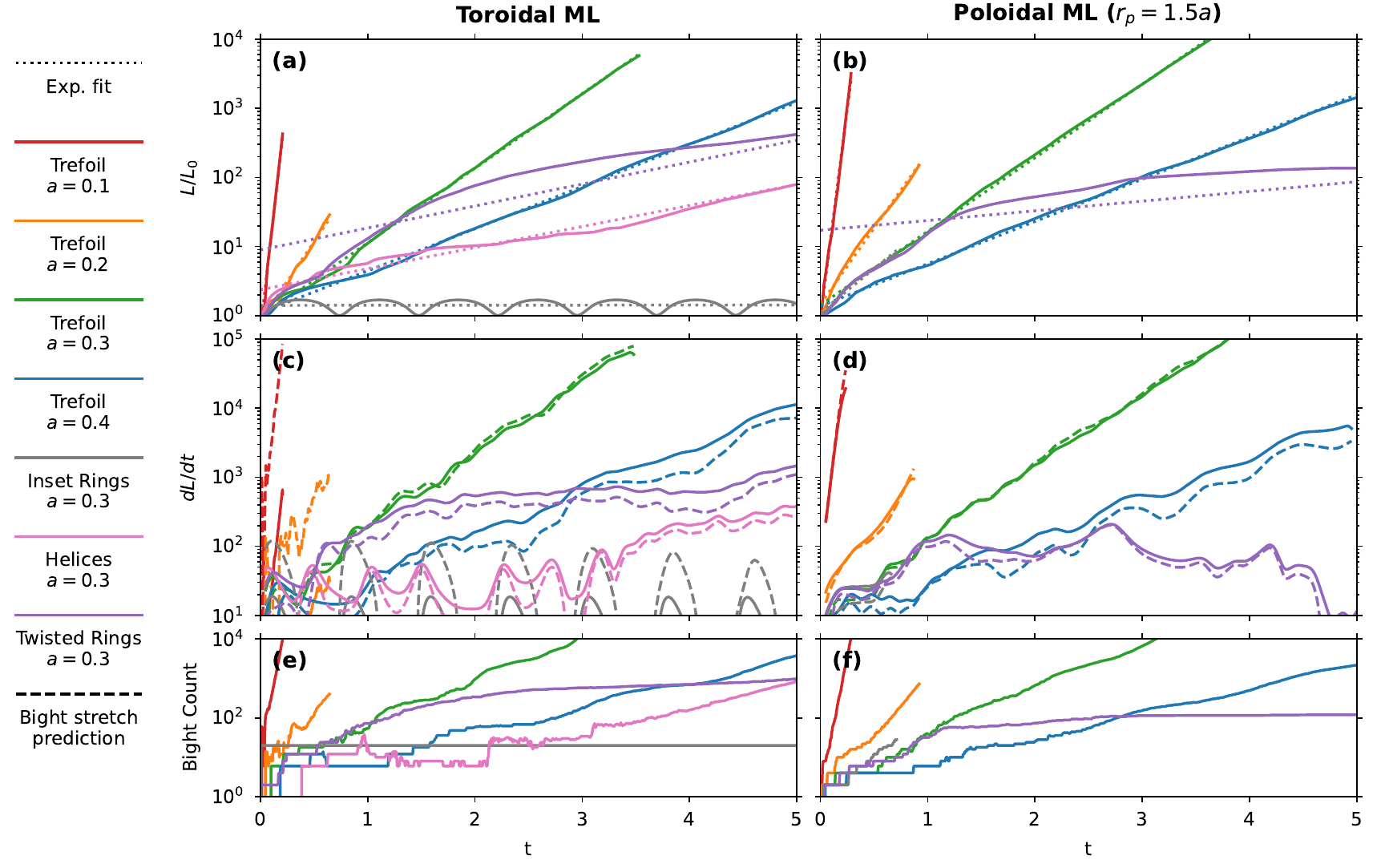}
\caption{
The stretching of material lines by vortices of varying shape and topology.
For all plots the left plots show inital toroidal material lines, and the right side shows poloidal material lines where $r_p = 1.5 a$.
\textbf{(a-b)} The total length as a function of time.  Dotted lines show exponential fits; note that in the cases of unknotted vortices it can be seen that the trend is not exponential, while it is for the trefoil knots.
\textbf{(c-d)} The stretching rate as a function of time.  The solid lines are computed from the numerical derivative of the length, with a Gaussian smoothing of width $\sigma_t = 0.05$ applied to remove high frequency noise.  The dashed lines shows an estimate of the stretch rate, obtained from \eqnref{bight-stretch}. 
\textbf{(e-f)} The bight count, obtained by locating the number of points on the material line where $\V u \cdot \Vh T$ changes sign.
\label{fig:stretch}
}
\end{figure*}

The stretching behavior for each vortex line is computed by advecting an initial material line which is a perfect circle oriented either toroidally (threaded through the center of the torus which defines the vortex lines), or poloidally (wrapping around the torus with radius $r = 1.5 a$).
The evolving length of all cases is shown in \figref{stretch}.

The simplest case we consider is a pair of inset circular vortices, which produce linear stretching analogous to the infinite straight line.
(Note that similar results can be obtained for a single vortex ring, which is not shown here because the pair of rings is a better comparison to the other cases.)


More dramatic cases of the stretching behavior are provided by the trefoil knots: in each case we see rapid exponential stretching of material lines, with an exponential rate which depends primarily on the aspect ratio.
To determine this rate, we fit the last 1.5 time units of each simulation to the equation:
\begin{align}
L(t) &= e^{a + k t},
\end{align}
where the values of $k$ obtained for each simulation are shown in \tableref{exp}.
The choice of material line also has a slight impact on the observed exponential stretching rate, although far less than the aspect ratio of the knot.

In each case, it can be seen that the evolving material line rapidly forms new bights when regions of it pass between the vortices.
These bights can be used to estimate the stretching rate (\figref{stretch}\textbf{(c-d)}); although it does not provide a quantitatively accurate result as in the case of the infinite line.
This is likely because new bights are continuously forming, thus one always expects to find bights that have not yet reached the long time limit of sharp tips.
Nonetheless it does indicate that the phenomenology of the stretching behavior is still explained by the bight picture.

The key distinction between the unknotted and knotted vortices is the ability to produce new bights continuously.
Indeed, for trefoil knots the number of bights is observed to grow exponentially, along with the length.
We note that if the probability of producing a new bight is proportional to the current length -- and the stretching rate is (roughly) proportional to the number of bights -- we should then obtain exponential growth.
Interestingly the twisted ring case does appear to have exponential growth (and a corresponding increase in number of bights) for short times.
Eventually, however, the material line is transported out of the bight forming region and the growth becomes linear.

The choice of torus knot aspect ratio affects the speed of stretching, but not the qualitative appearance of the stretching material line.
The increase in rate can be explained by the fact that the time it takes for a point to be transported around the vortices will be much reduced for smaller aspect ratio.

To explore which features of the vortex knot produce the transition between linear and exponential stretching, we also distort the unknotted rings in a manner analogous to the knots.
The simplest way to do this is to make each of the rings helices, so that the vortex lines have a torsion comparable to the knotted case.
However, these helices produces linear stretching, as with the undistorted rings.

A more interesting case is provided by twisting each of the rings around the surface of a torus, which can be done while still leaving them unlinked and unknotted (i.e.~with trivial toplogy in the sense of knot theory).
This produces vortices with local sections that look nearly identical to the knot.
Interestingly, for short 
periods of time this configuration appears to produce exponential stretching, corresponding to periods of rapid bight production.
On longer timescales the stretching becomes linear, and the bight production slows.

Thus, we observe that the twisted rings do indeed have a region which is capable of producing new bights and providing stretching behavior analogous to the knotted case.
However, this behavior can not be maintained, as the material line is eventually transported out of the region of bight formation where it remains at long time scales.

\begin{figure}
\includegraphics{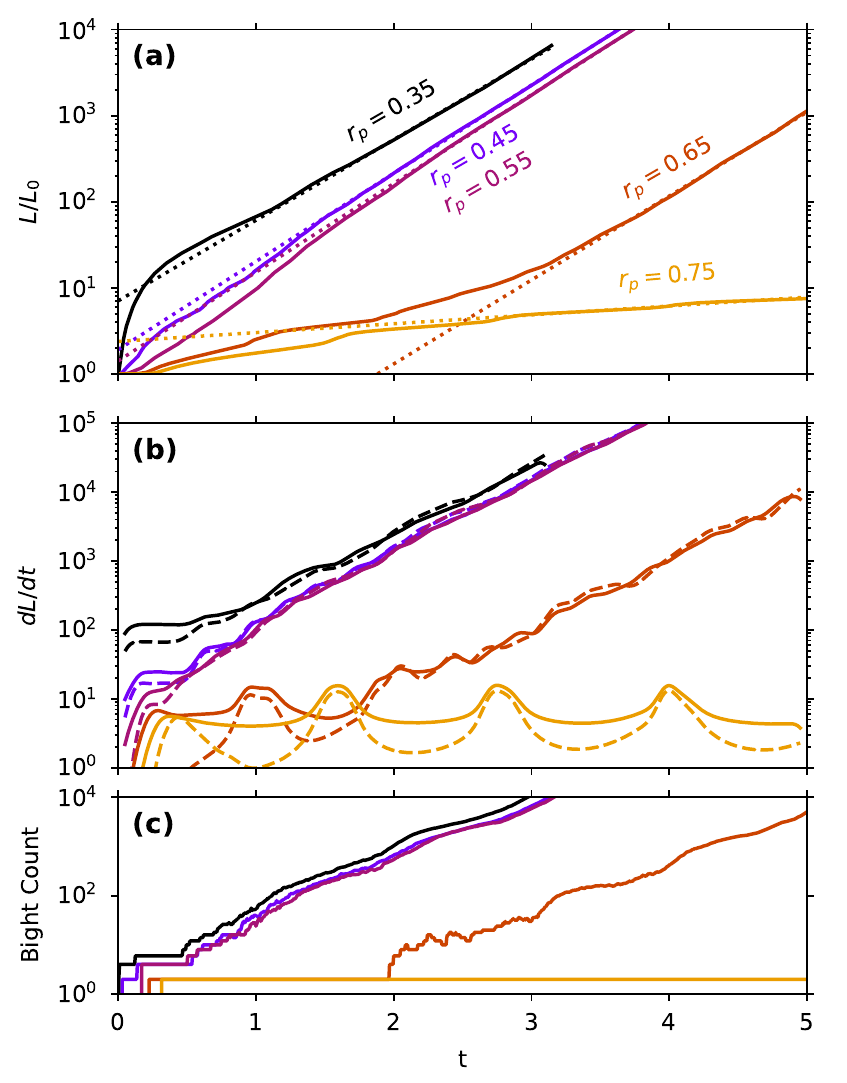}
\caption{
Stretching of polodial material lines of varying radius, $r_p$ for a trefoil knot of aspect ratio $a=0.3$.
For $r_p = 0.75$ new bights are never formed and the material line has a limited stretching rate; in all other cases new bights are formed and exponential stretching is reached with approximately the same growth rate.
See \figref{ftle}\textbf{c} for an overlay of the polloidal material lines with the FTLE values.
\label{fig:pol-stretch}
}
\end{figure}

One might expect that the stretching speed could be modified by changing the radius of the initially poloidal material line, $r_p$.
We conducted an additional 5 simulations for a torus knot with aspect ratio, $a=0.3$, and material line radius, $r_p = 0.35 - 0.75$.
The stretching behavior is shown in \figref{pol-stretch}.
We observe that the exponential stretching rate for $r_p = 0.35 - 0.65$ is nearly the same, ranging from $k = 2.17-2.37$.
For $r_p = 0.75$ we observe no bight production and linear growth of the material line.
Evidently if we are sufficiently far from the vortex the bight production is suppressed; the reason for this is discussed in the next section.

Might it be possible to produce topologically trivial shapes with continuous bight production?
This seems unlikely for the case of the vortex lines confined to surface of a torus: if they are continuously twisted in one direction the resulting shapes with be topologically linked or knotted.
If the twist rate oscillates -- as in the twisted ring case -- it will always have locally untwisted regions which seem not to produce exponential stretching.
If the material lines are always transported into these regions, they will ultimately only produce linear stretching.
It may be possible to produce shapes \emph{not} confined to the surface of a torus that do not have this limitation, which we leave as an avenue for future studies.

\begin{figure*}
\includegraphics{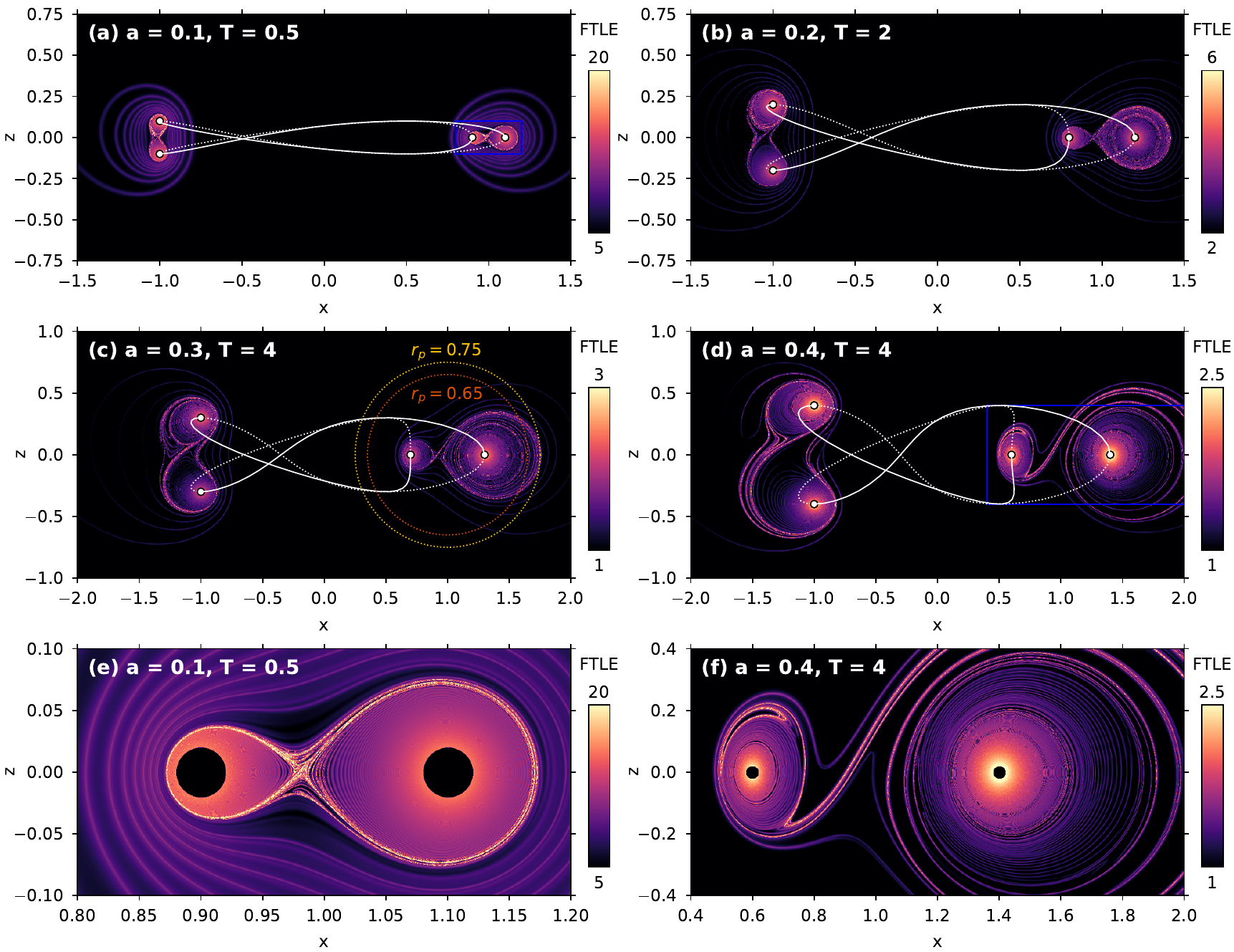}
\caption{
A map of finite time Lyapunov exponents, computed for trefoil torus knots of varying aspect ratio.  
The exponents are computed for a time interval of $T=30$, and values are shown for a slice in the $y=0$ plane.
The white (dashed) lines show the location of the vortex lines in front of (or behind) the plotted plane.
The first four panels \textbf{(a-d)} show both sides of the vortex for aspect ratios from $a=$0.1--0.4, while \text{(e-f)} shows a zoomed in portion of two cases.
(The zoomed in regions are indicated with blue boxes in \textbf{(a)} and \textbf{(d)})
The FTLE isn't computed for points within a radius of $r=0.02$ of the vortex cores (black circles) to prevent numerical precision issues.
Poloidal material line locations for $r_p = 0.65, 0.75$ are indicated in \textbf{(c)}.
Note that the $r_p = 0.75$ line lies just outside the unstable manifold; as shown in \figref{pol-stretch} this material line has only linear stretching.
\label{fig:ftle}
}
\end{figure*}

\section{Finite Time Lyapunov exponents}
\label{sec:ftle}

An alternative method to probe for exponential stretching is to compute the finite time Lyapunov exponent (FTLE) of the flow field \cite{hallerLagrangianCoherentStructures2015}.
The FTLE characterizes the exponential stretching rates of parcels of fluid in the direction of maximum strain.

We compute the FTLE using the Cauchy strain tensor; for an initial point, $\V x(t=0)$, this is given by:
\begin{align}
    C_{ij}(T) &= \sum_k \pd{x_k(T)}{x_i(0)} \pd{x_k(T)}{x_j(0)},
\end{align}
where $\V x(T)$ is the point $\V x(0)$ transported forward in time an interval $T$, and in practice the derivatives are computed by attaching infinitesimal unit vectors to each point (see \appendixref{ftle-comp}).

The FTLE is then given by:
\begin{equation}
    \Lambda_T (\V x) = \frac{1}{T} \log \sqrt{\lambda_{max} (\V x)},
\end{equation}
where $\lambda_{max}(\V x)$ is the largest eigenvalue of the deformation tensor, $C_{ij}$, computed for an \emph{initial} position $\V x$.

FTLE maps for knotted vortices of varying aspect ratio are shown in \figref{ftle}.
In each case, the regions of highest exponential growth appear in the regions \emph{between} the vortices.
This is despite the fact that the regions of highest instantaneous shear are immediately adjacent to the vortices; these regions do not produce new bights, and so produce only linear stretching in the long time limit.
(If the FTLE is computed for larger $T$, the linearly stretching regions near the core will have a smaller values, but limits of numerical precision prevent us from accurately computing the exponent in these regions; the simulations are stopped when the longest stretched unit vectors are stretched by a factor of $\sim 10^{10}$.)

Most notably, the FTLE plots reveal ridges of high stretching -- unstable manifolds -- which connect sections of the vortex (e.g.~in the center of \figref{ftle}\textbf{(e-f)}.
These appear to be the regions responsible for producing new bights; we note that if the material line does not cross one of these ridges we do not observe exponential stretching.
For example, for $a=0.3$, $r_p =0.75$, we observe only linear growth (\figref{pol-stretch}).
This material line lies just outside the FTLE ridge; for a slightly smaller value of $r_p = 0.65$ which \emph{does} cross the ridge we do observe exponential growth after an initial linear period.
Additionally, we note that the quantitative rate of exponential stretching observed for both toroidal and poloidal material lines has roughly the same time constant as the peak value on the ridges of high stretching observed in the FTLE plots.

\section{Conclusion and Discussion}
In this manuscript we have compared the stretching of arbitrary material lines by vortical flow fields of varying topology and geometry.
In particular we have shown that the flow fields of knotted vortices produce exponential stretching of material lines in certain regions, which is not seen in the stretching of flow fields of unknotted vortices.
This change in the qualitative character of these flow fields is confirmed by computing the FTLE of the flow fields.
The FTLE results demonstrate that it is the regions \emph{between} the vortices which are responsible for exponential stretching.
Intuitively, this can be explained by the production of `bights' in these regions.
Conversely when a region of flow is dominated by a \emph{single} vortex the rapid strain prevents the formation of new bights, ultimately producing less stretching over the long term even though the local strain rate may be much higher.

These results suggest previously unknown connections between previous results in the physics of mixing and vortex dominated fluid flows.
In 2D flows, it is well known that the topology of a time-dependent flow fields is connected to long term exponential stretching \cite{boylandTopologicalFluidMechanics2000,thiffeaultTopologyBraidsMixing2006}.
In this case, exponential stretching is produced by a `stretching and folding' action reminiscent of a taffy puller.
Our work suggests that topology in space alone can play a similar role in 3D flows; in our case the `folding' of the material lines is indicated by the production of bights.

These results may also offer an explanation for the apparent instability of knotted vortices.
It has previously been observed -- in both experiments and simulations  -- that linked and knotted vortices are highly unstable to self-stretching \cite{klecknerHowSuperfluidVortex2016,binyshStableUnstableVortex2019, kimuraScalingPropertiesVortex2018, klecknerCreationDynamicsKnotted2013, klecknerLifeVortexKnot2014, kerrTrefoilKnotTimescales2018, promentTorusQuantumVortex2014}.
Although our results do not model the \emph{self-stretching} of vortex lines (indeed, we treat the vortices as fixed), they do demonstrate that vortex topology has a dramatic effect on the stretching behavior of the a flow field.
As vortex lines themselves are transported by flows, it follows that we would also expect a qualitative difference in their evolution.

Although in principle one could model the stretching produced by vortices whose shape evolves in time, as previously noted knotted vortices will rapidly approach reconnection events.
Thus, a dynamic model would be limited in the total amount of time which the vortices remain knotted, making it difficult or impossible to separate exponential from non-exponential stretching.
Interestingly, our results suggest that a more complex model is not required: the advection of material lines by a `frozen' flow field may be sufficient to indicate their long term stability.
This also suggests that the non-linearities present in the full Navier-Stokes equation are not required to explain the difference in stability between knotted and unkontted vortices.


These results have potential relevance for a wide variety of vortex-dominated flows, such as tornadoes, flows around aerodynamic surfaces, and turbulence. 
For example, we note that enstrophy production -- i.e.~vortex stretching -- is one of the hallmarks of 3D turbulence, and is one of the features which distinguishes 2D from 3D flows \cite{ouelletteTurbulenceTwoDimensions2012}.
Although further work is required to show that these results hold for a wider variety of flow configurations -- e.g.~multiple vortices of different circulation, those with finite core size, etc. -- it does suggest the particular shape the `tangled' vortices in a turbulent flow may offer clues as to when and why vortex stretching occurs.

Finally, we note that our work does not rule out the possibility that that some complex unkotted configuration of vortex lines may produce exponential stretching, or that \emph{all} knotted vortex fields do.
In future work it may be possible to prove this using newly developed 3D analogues of the techniques used in 2D topological mixing \cite{smithTopologicalChaosThreedimensional2017}.
Nonetheless, the current results demonstrate a clear connection between vortex topology and stretching behavior mediated by bight production.

\appendix

\section{Infinite Straight Line Vortex}
\subsection{Global vs.~Local Stretching}
\label{sec:global-local}

\Sectionref{inf-bights} describes stretching of material lines using a curvature based formula.
This formula -- which is only valid for \emph{closed} material lines -- attributes the stretching to the curved regions of the vortex.
Naturally, a straight line segment in a varying flow field can also stretch; the equivalent formula for an open material line, which includes end terms, is:
\begin{align}
    \dot L &= -\int_{0}^{L} \V u \cdot \kappa \Vh N\ ds + \V u_f \cdot \Vh T_f - \V u_i \cdot \Vh T_i, 
\end{align}
where $\V u_{i/f}$ is the velocity at the initial/final point along the curve, and $\V T_{i/f}$ is the corresponding tangent vector.
In the case of a straight line, the stretching is simply given by the differential velocity at the two end points, which function in a manner similar to bights.

This formulation -- for either open or closed curves -- does not indicate \emph{where} along the line the stretching actually occurs.
(For example, a straight line in a varying flow field will have local stretching along its length; this is not predicted by the curvature formulation, even though the total length change is.)
If required, local stretching can be obtained using the strain tensor:
\begin{align}
\frac{\dot{\ell}}{\ell} &= \del_{\Vh T} \V u \cdot \Vh T,
\end{align}
Calculating this quantity for a material line near an infinite vortex shows that the stretching also occurs in the region between the bights, despite the fact that $\V u \cdot \kappa \Vh N \approx 0$ here.
Nonetheless, one can treat the bights (or line ends) as responsible for the stretching, and analyzing the stretching in terms of these points provides a simple, intuitive picture of the dynamics.
In other words: the bight/curvature picture does not indicate \emph{where} stretching of the material line will happen, but does indicate \emph{when} it will happen.

\subsection{Alignment with Flow Direction}
Consider a short material line embedded in the flow of an infinite vortex along the $z$-axis.
Assuming a circulation $\Gamma = 2 \pi$, the flow field in cylindrical coordinates is:
\begin{align}
	\V u = \frac{\Vh \phi}{\rho}
\end{align}
Where $\rho$ is the distance from the $z$-axis.
The transport of a single point is given by:
\begin{align}
	\phi(t) &= \phi(0) + \frac{t}{\rho(0)^2}
\end{align}
With the other coordinates ($\rho$ and $z$) remaining constant.


If our short material line segment is described by tangent vector $\V T$ (where $|\V T|$ is identified with the infinitesimal segment length), we can derive its time evolution:
\begin{align}
	\V T(t) &=  T_\rho(0) \Vh \rho + \left[T_\phi(0) - \frac{2 t T_\rho(0)}{\rho(0)^2}\right] \Vh \phi + T_z(0) \Vh z
\end{align}
Note that these unit vectors are specified at the location of the moving point; if expressed in rectangular unit vectors, an overall rotation would also need to be included.
From this equation, we can see that the material line tangent vector will simply increase or decrease in the $\phi$ direction, depending on the sign of $T_\rho$.
In other words, segments pointing out from the $z$-axis will asymptotically align with $-\Vh \phi$ (or $-\V u$), and segments pointing in with $+\Vh \phi$ (or $+\V u$).
Furthermore, we can see that this segment stretches, and that as $t\rightarrow \infty$ we obtain $|\V T| \propto t$.
This implies that \emph{any} material line which stretch at most linearly in the long time limit.

\section{Simulation Details}
\label{sec:sim-details}
\subsection{Flow Field}
We define our vortex path as a series of polygons with points $\V r_j$.
The flow field for a polygonal vortex can be exactly computed as \cite{hansonCompactExpressionsBiot2002}: 
\begin{align}
    \label{eqn:poly_flow}
    \V u(\V x) &= \sum_j \frac{\Gamma}{4\pi}\ \frac{2 \epsilon_j}{1 - \epsilon_j^2}\ \frac{\Vh \Delta_j \times \V R_j}{R_j R_{j+1}}\\
    \V R_j &= \V r_j - \V x;&R_j \equiv |\V R_j|\\
    \V \Delta_j &= \V r_j - \V r_{j+1};& \Vh \Delta_j \equiv \frac{\V \Delta_j}{|\V \Delta_j|}\\
    \epsilon_j &= \frac{|\V r_j - \V r_{j+1}|}{r_j + r_{j+1}},
\end{align}
where the sum is over all polygon edges.

To model material lines or compute FTLE fields, we track individual points in the field, $\V x_i$, with one or more attached infinitesimal vectors, $\V V_{i, n}$.
These points are advected in the flow field, i.e.:
\begin{align}
    \pd{\V x_i}{t} &= \V u\left[\V x_i(t)\right] \label{eqn:dx}\\
    \pd{\V V_{i, n}}{t} &= \del_{\V V_{i, n}} \V u\left[\V x_i(t)\right], \label{eqn:dv}
\end{align}
where an exact expression for the gradient of the flow field is given by:
\begin{align}
    \nonumber  \del_{\V v} \V u &= \sum_j \frac{\Gamma}{4\pi} \  \frac{2 \epsilon}{1 - \epsilon^2}\ \frac{\Vh \Delta_j}{r_i r_f} \times\\
& \left[\V v - \V r_i \left(\V v \cdot \left[\frac{\Vh r_i}{r_i} + \frac{\Vh r_f}{r_f} + \frac{\Vh r_i + \Vh r_f}{r_i r_f}\ \frac{1 + \epsilon^2}{1 - \epsilon^2} \right] \right)\right]
\label{eqn:vecflow}
\end{align}
Numerically, both are integrated using the Dormand-Prince method \cite{dormandFamilyEmbeddedRungeKutta1980, shampinePracticalRungeKuttaFormulas1986} with an absolute velocity tolerance of $10^{-8}$.

\begin{table*}
\begin{tabular} { c|c|c|c|c } 
 \textbf{Vortex Shape} & \textbf{Aspect Ratio} & \textbf{Tor.~Interval} & \textbf{Pol.~ML Radius} & \textbf{Pol.~Interval}\\
 \hline
& $a=0.1$ & $T_{resample} = 4\cdot 10^{-6}$ & $r_p = 0.15$ & $T_{resample} = 10^{-5}$\\
\cline{2-5}
 & $a=0.2$ & $T_{resample} = 1.6\cdot 10^{-5}$ & $r_p = 0.3$ & $T_{resample} = 5\cdot 10^{-5}$\\
\cline{2-5}
Trefoil Knots &$a=0.3$ & $T_{resample} = 3.6\cdot 10^{-5}$ & $r_p = 0.35$ & $T_{resample} = 5\cdot 10^{-5}$  \\ 
 &&& $r_p = 0.45-0.75$ & $T_{resample} = 10^{-4}$\\ 
 \cline{2-5}
& $a=0.4$ & $T_{resample} = 6.4\cdot 10^{-5}$  & $r_p = 0.6$ & $T_{resample} = 10^{-4}$\\
 \hline
All Unkots & $a=0.3$ & $T_{resample} = 10^{-4}$ & $r_p=0.45$ & $T_{resample} = 10^{-4}$
\end{tabular}
\caption{
Resampling intervals used for various simulations.
Note that the interval must be reduced for simulations where a material line comes close to the vortex.
This is needed to ensure that the it is resampled roughly the same number of times per transit around the vortex: in practice this means the interval scales as $T_{resample} \sim a^{-2}$.
In each case the simulations were manually checked to ensure the sampling remained qualitatively accurate -- insufficient resampling generally produces a numerically unstable result.
}
\label{table:resample}
\end{table*}

\subsection{B\'ezier Curves}
Because some sections of the material line will stretch far more than others, to accurately track its length we need a method for inserting new points into the curve as it evolves.
To implement this, we approximate the material line as a series of cubic B\'ezier curves.
We do this by representing the material line as a series of points along the curve, $\V x_i$, with a single attached tangent vector, $\V V_i$.
These points and vectors evolve according to \eqnref{dx} and \eqnref{dv}.
Each point also has a attached segment length, $s_i$, which is constant in time excepting periodic resampling of the curve (see below).

The implicit cubic B\'ezier curve for each segment is then  given by:
\begin{align}
&\nonumber \V x_i(z) = (1-z)^3 \vec{\bf P}_0 + 3(1-z)^2 z \vec{\bf P}_1 + \\
&\quad \quad 3(1-z) z^2 \vec{\bf P}_2  + z^3 \vec{\bf P}_3
\end{align}
where $z=0-1$ is the position along each curved segment and the control points are define as:
\begin{align}
    \vec{\bf P}_0 &= \V{x}_i\\
    \vec{\bf P}_1 &= \V{x}_i + \frac{s_i}{3} \V{V}_i\\
    \vec{\bf P}_2 &= \V{x}_{i+1} - \frac{s_i}{3} \V{V}_{i+1} \label{eqn:P2}\\
    \vec{\bf P}_3 &= \V{x}_{i+1}.
\end{align}
Note that $P_2$ depends on $s_i$, not $s_{i+1}$, even though the tangent vector is given by $\V V_{i+1}$.
Initially, $s_i$ corresponds to the arc length of this segment, and the tangent vectors, $\V V_i$, are all unit length.
As the curve is advected this will result in $\V V_i$ changing length to compensate for stretching, rather than including it directly in $s_i$.
This is done to ensure that $\pd{x_i}{s}(z=0) = \V V_i$ and  $\pd{x_i}{s}(z=1) = \V V_{i+1}$.

At regular intervals, the path is resampled: $s_i$ are replaced with the estimated length of each Bezi\'er segment, and the tangent vectors are re-normalized.
The actual length of each curve is estimated by sampling each curve at 100 points and computing the total distance between these points.
See \tableref{resample} for the resampling interval used for various simulations.

\subsection{Addition and Removal of Points}
An angular and length error is considered when decided to insert new points into the curves which define the material lines.
A error number for each segment, $N_i$,  is computed using:
\begin{align}
    N_i &= \textrm{max} \left[\frac{| \Vh V_i \times \Vh \Delta_i | + | \Vh V_{i+1} \times \Vh \Delta_i |}{\epsilon_a}, \frac{|\V \Delta_i|}{\epsilon_\ell}\right],
\end{align}
where $\V \Delta_i = \V x_{i+1} - \V x_i$ is the segment displacement vector ($\Vh \Delta$ is the normalized equivalent), $\epsilon_a = 0.1$ is the angular tolerance parameter, and $\epsilon_\ell = 0.1$ is the length tolerance parameter.

The addition or removal of points happens at the same interval as the resampling described above.
If $N_i > 1$ for any segment, new points are inserted between the ends of the segment using the Bezier approximation; the number of inserted points is $\textrm{ceil}[N_i]- 1$.
To remove redundant points, we consider pairs of segments with even and odd indices; if the sum of $N_i$ for these neighboring segments is $< 0.9$, the midpoint is removed.

\subsection{FTLE Computation}
\label{sec:ftle-comp}
To compute the FTLE of vortical flow fields, we model the advection of a point, $\V x(t)$, with three attached infinitesimal vectors, $\V V_n(t)$.
These vectors are initialized so that $\V V_n(t=0) = \Vh e_n$, where $\Vh e_n$ are the Cartesian unit vectors.
In this representation the strain tensor and Cauchy strain tensor are given by:
\begin{align}
    F_{ij} &= \Vh e_i \cdot \V V_j\\
    C_{ij} &=  \sum_k(\V V_i \cdot \Vh e_k)(\V V_j \cdot \Vh e_k).
\end{align}

In practice, the stretching can be so rapid that numerical precision can limit the maximum observed FTLE value.
To limit this, we show FTLE values computed for a time when the relative stretching of all vectors is less than $10^{10}$.

\subsection{Simulation Parameters}
\label{sec:sim-params}
All vortex paths used in these simulations are defined on the surface of a torus; the conversion between torroidal ($R$, $a$, $\phi$, and $\alpha$), cylindrical ($\rho$, $\phi$, and $z$), and cartesian ($x$, $y$, and $z$) coordinates is given by:
\begin{align}
    \rho &= R + a \cos \alpha\\
    x &= \rho \cos \phi\\
    y &= \rho \sin \phi\\
    z &= a \sin \alpha,
\end{align}
where $\phi$ is azimuthal angle, $R$/$a$ are the major/minor radius of the torus and $\alpha$ is the polloidal angle.
The aspect ratio of the torus is given by $a/R$; for most cases $R=1$, so the aspect ratio is given by $a$.

Parametric equations for the vortex knots are given by:
\begin{align}
    \textrm{Trefoil knot: } \phi &= (0-4\pi)\\
    \alpha &= \frac{3}{2} \phi\\
    R &= 1,
\end{align}
where $a = (0.1, 0.2, 0.3, 0.4)$ is the aspect ratio, which varies for different simulations.

Note that this knotted solution wraps around the torus twice in the azimuthal direction; we desire for our unknotted cases to do the same for a more direct comparison.
Since one can not define a closed single path on the surface of a torus which wraps twice and is itself unknotted,  the three unknotted cases are each composed of two separated vortices which each wrap once around a torus.
The parametric equations for these are:
\begin{align}
    \textrm{Inset rings: } \phi_{1/2} &= (0 - 2 \pi)\\
    a_{1/2} &= 0\\
    R_1 &= 0.7\\
    R_2 &= 1.3
\end{align}
\begin{align}
    \textrm{Inset Helices: } \phi_{1/2} &= (0 - 2 \pi)\\
    \alpha_{1/2} &= 3 \phi_{1/2}\\
    a_{1/2} &= 0.3\\
    R_1 &= 0.85\\
    R_2 &= 1.15
\end{align}
\begin{align}
    \textrm{Twisted Rings: }  \phi_{1/2} &= (0 - 2 \pi)\\
    \alpha_1 &= \frac{3}{2} \sin \phi_1\\
    \alpha_2 &= \pi + \frac{3}{2} \sin \phi_2\\
    R_{1/2} &= 1\\
    a_{1/2} &= 0.3,
\end{align}
where in each case we have chosen the aspect ratios and/or separation between vortices to match the $a=0.3$ torus knot case.
Additionally, the maximum twist angle in the twisted ring case is chosen to match the (constant) twist angle in the trefoil knot case.

\bibliography{refs}

\end{document}